\documentclass[twocolumn,showpacs,preprintnumbers,amsmath,amssymb,superscriptaddress]{revtex4}

\usepackage{graphicx}
\usepackage{dcolumn}
\usepackage{bm}
\usepackage{epsfig}

\begin{document}

\preprint{draft}

\title{Suppression of the stellar enhancement factor and the reaction $^{85}$Rb(p,n)$^{85}$Sr}

\author{T.\,Rauscher}%
\email{Thomas.Rauscher@unibas.ch}
\affiliation{%
Department of Physics, University of Basel, CH-4056 Basel, Switzerland}%
\author{G. G.\,Kiss}%
\author{Gy.\,Gy\"urky}%
\author{A.\,Simon}%
\author{Zs.\,F\"ul\"op}%
\author{E.\,Somorjai}%
\affiliation{%
Institute of Nuclear Research (ATOMKI), H-4001 Debrecen, POB.51., Hungary}%

\date{\today}

\begin{abstract}
It is shown that a Coulomb suppression of the stellar enhancement factor occurs in many
endothermic reactions at and far from stability. Contrary to common assumptions,
reaction measurements for astrophysics with minimal impact of stellar enhancement should be
preferably performed for those reactions instead of their reverses, despite of their
negative $Q$ value.
As a demonstration, the cross section of the astrophysically relevant $^{85}$Rb(p,n)$^{85}$Sr
reaction has been measured by activation
between $2.16\leq E_\mathrm{c.m.}\leq 3.96$ MeV and 
the astrophysical reaction rates at $p$ process temperatures for (p,n)
as well as (n,p) are directly inferred from the data. Additionally, our results confirm a previously derived modification of a global optical proton potential. The presented
arguments are also relevant for other $\alpha$- and proton-induced reactions in the $p$, $rp$, and $\nu p$ processes.
\end{abstract}

\pacs{26.50.+x Nuclear physics aspects of novae, supernovae, and other explosive environments, 24.60.Dr Statistical compound-nucleus reactions, 27.50.+e 59 $\leq$ A $\leq$ 89 }%

\maketitle

\section{Introduction}
\label{sec:intro}

Astrophysical reaction rates are central to tracing changes in the abundances of nuclei by nuclear reactions. They provide the temperature- and density-dependent coefficients entering reaction networks, the large sets of coupled differential equations required to study nucleosynthesis and energy generation in astrophysical environments. The reaction rates are computed from reaction cross sections which, in turn, may be predicted in theoretical models or extracted from experiments. In addition to the difficulties arising in the determination of the cross sections, the conversion to reaction rates is further complicated by modifications of the rates in a hot plasma and the fact that the rates of forward and reverse rate for the same reaction have to be consistent to ensure numerical stability and proper equilibrium abundances at high temperature. Both issues can be addressed at once by accounting for the thermal population of target states which leads to \textit{stellar} rates obeying a reciprocity relation between forward and reverse rate. Using this reciprocity, knowledge of the rate in only one direction is needed because the other reaction direction can be directly computed from that rate, thus ensuring consistency.

For numerical reasons, further elaborated in Sec.\ \ref{sec:sef}, it is usually preferable to start from the rate of a reaction with positive reaction $Q$ value when computing the rate for its inverse reaction. Even more importantly, experimentalists want to determine rates as close as possible to the actual stellar rates, i.e.\ rates with minimal thermal population effects of the target. Again, it can be argued that this is the case for exothermic reactions. This led to the commonly applied rule that measurements of exothermic reactions are more important than those of endothermic ones. In this paper we show that there is a considerable number of reactions for which a suppression effect brings the stellar rate of an endothermic reaction closer to the laboratory value than its exothermic counterpart.

As an example of how to exploit this suppression effect and to obtain stellar rates from a measurement of an endothermic reaction, we experimentally studied the reaction $^{85}$Rb(p,n)$^{85}$Sr, having $Q=-1.847$ MeV. The importance of the reaction is manifold.
In the last several years a number of proton capture cross section measurements with
relevance for $\gamma$ process studies have been carried out 
(see, e.g., \cite{kiss07} and references therein). The $\gamma$ process was shown to synthesize
$p$ nuclides (proton-rich isotopes not accessible to the $s$ and $r$ processes) by a series of photodisintegrations of stable nuclides
in hot layers of massive stars \cite{woo78,arn03,rau02,rau06}.
Recently, systematic
$\gamma$ process simulations found not only that photodisintegration reactions
are important but also that (p,n) reactions, and in particular $^{85}$Rb(p,n)$^{85}$Sr,
strongly influence the final $p$ abundances \cite{rap06}. Additionally, this
reaction is well suited to test
the optical potential used for calculating the
interaction between protons and target nuclei.

We commence by outlining the theoretical background regarding stellar rates and the suppression effect in Secs.\ \ref{sec:ratesintro} and \ref{sec:sef}. The results of a large-scale study of the effect in the full extension of the nuclear chart are discussed in Sec.\ \ref{sec:theoryresults}. Focusing on the reaction $^{85}$Rb(p,n)$^{85}$Sr, their relevance is discussed in Sec.\ \ref{sec:exp}, the experimental details are provided in Secs.\ \ref{sec:exptarget}--\ref{sec:activity}, and the astrophysically relevant rates are derived in Sec.\ \ref{sec:astrorate}. Additionally, Sec.\ \ref{sec:potential} discusses implications of our new experimental results for the proton optical potential. Finally, a summary is given in Sec.\ \ref{sec:summary}.

A brief account of our findings was already given in \cite{kiss08}. The present follow-up paper expands the discussion and also provides additional results in all parts of this investigation.

\section{Suppression of the stellar enhancement}

\subsection{Stellar reaction rates}
\label{sec:ratesintro}
The stellar enhancement factor (SEF) $f$ is defined as the ratio of the stellar rate $r^*$ relative to the ground state rate $r^\mathrm{g.s.}$ \cite{ilibook}
\begin{equation}
\label{eq:sef}
f=\frac{r^*}{r^\mathrm{g.s.}}=\frac{r^*}{r^\mathrm{lab}} \quad.
\end{equation}
The rate $r^\mathrm{lab}$ derived from cross sections $\sigma^\mathrm{lab}$ measured in the laboratory is the same as $r^\mathrm{g.s.}$ because so far all experiments use target nuclei in their ground states. The SEF is a measure of the influence of the thermally excited target states in the hot plasma.
\begin{figure}
\resizebox{\columnwidth}{!}{\rotatebox{0}{\includegraphics[clip=]{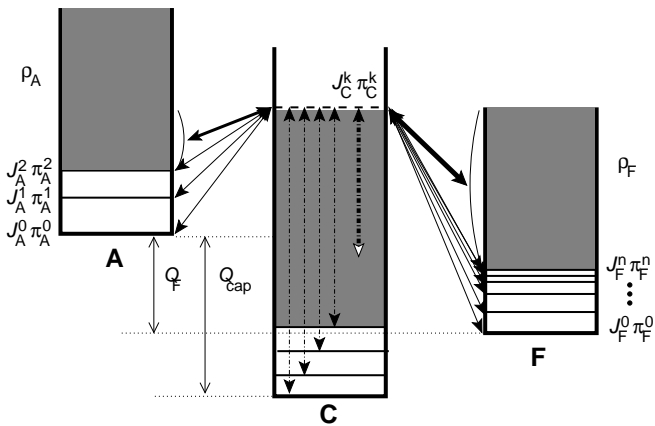}}}
\caption{Schematic view of the transitions (full arrows denote particle transitions, dashed arrows are $\gamma$ transitions) in a compound reaction involving the nuclei A and F, and proceeding via a compound state (horizontal dashed line) with spin $J_\mathrm{C}^k$ and parity $\pi_\mathrm{C}^k$ in the compound nucleus C. The reaction $Q$ values for the capture reaction (Q$_\mathrm{cap}$) and the reaction A$\rightarrow$F (Q$_\mathrm{F}$) are given by the mass differences of the involved nuclei. The effective cross section $\sigma^\mathrm{eff}$ (Eqs.\ \ref{eq:simprate} and \ref{eq:effcs}) for a reaction type is a sum over all energetically possible transitions to bound states (capture: in nuclei A and C; otherwise: in nuclei A and F) from the compound level as shown here (see text for details). In each nucleus, a number of low-lying states with given spin $J$ and parity $\pi$ is explicitely specified. Above the last state, transitions can be computed by integrating over nuclear level densities (shaded areas). In stellar cross sections $\sigma^*$ all transitions are additionally weighted by a Boltzmann distribution factor depending on the stellar temperature, spin, and the excitation energy of the involved state (see Eq.\ \ref{eq:pop}).\label{fig:eff}
}
\end{figure}

Astrophysical reaction rates are usually defined as giving the number of a specific reaction occurring per time. Here, we constrain ourselves to two-body reactions of nuclei and nucleons. The concept of the stellar rates suppression introduced below is easily extended to other reaction types.
Reaction cross sections are folded with the energy distribution of the interacting nuclei to obtain the reaction rate. The energy distributions of nuclei and nucleons in an astrophysical plasma follow Maxwell-Boltzmann distributions in most applications, thus yielding \cite{ilibook}
\begin{eqnarray}
\label{eq:rate}
r_i&=&\frac{n_1 n_2}{1+\delta_{12}} \frac{F}{\left( kT\right) ^{3/2}} \int_0 ^\infty \sigma_i E e^{-\frac{E}{kT}}\,dE \\
&=&\frac{n_1 n_2}{1+\delta_{12}} \mathcal{R}_i \nonumber
\end{eqnarray}
for reactions proceeding from target state $i$ with reaction cross section $\sigma_i$, where $n_1$, $n_2$ are the number densities of the interacting nuclei, $T$ is the plasma temperature, $k$ denotes the Boltzmann constant, and $F$ is a renormalization factor $F=\sqrt{8/(\pi \mu)}$ with $\mu=A_1 A_2/\left( A_1+A_2\right)$ being the reduced mass number $A$.

When nuclei are in thermal equilibrium with their environment, their excited states are populated according to a Boltzmann factor \cite{ilibook}
\begin{equation}
\label{eq:pop}
P_i=\frac{\left( 2J_i+1 \right) e^{-\frac{E_i}{kT}}}{\sum_n {\left(2J_n+1\right) e^{-\frac{E_n}{kT}}}} \quad,
\end{equation}
with $P_i$, $J_i$, $E_i$ denoting the relative population, spin, and excitation energy of state $i$, respectively. Each of the states is bombarded with Maxwell-Boltzmann distributed projectiles which would require to have a separate rate for each target state weighted by the population factor of the state $i$ from which the reaction proceeds. It was shown in \cite{fow74} (see also \cite{hol76}) that by making use of the reciprocity theorem for nuclear reactions and detailed balance (assuming thermalization of both initial and final states of a reaction), for compound reactions the rate equation can be simplified to
\begin{eqnarray}
\label{eq:simprate}
\mathcal{R}^*&=&\frac{F}{\left( kT\right) ^{3/2}} \sum_i \left( \int_0^\infty P_i (T) \sigma_i\left(E^i\right) E^i e^{-\frac{E^i}{kT}} \,dE^i \right) \nonumber \\
&=&\frac{\left( 2J^0+1\right) F}{\left( kT\right) ^{3/2} G} \int_0 ^\infty \sigma^\mathrm{eff} \left(E\right) E e^{-\frac{E}{kT}}\,dE \quad, \label{eq:reactivity} \\
r^*&=&\frac{n_1n_2}{1+\delta_{12}}\mathcal{R}^* \quad .
\end{eqnarray}
In order to avoid additional computations caused by the population coefficients and also to avoid having a temperature dependent stellar cross section, the effective cross section $\sigma^\mathrm{eff}$ was introduced above, which sums over all bound states in the initial \textit{and} final system (denoted by $i$ and $j$, respectively; the
energetics of the transitions is shown in Fig.\ \ref{fig:eff}) \cite{hol76}:
\begin{equation}
\label{eq:effcs}
\sigma^\mathrm{eff}=\sum_i \sum_j \sigma_{ij} \quad.
\end{equation}
This is a theoretical construct (as any measurement would always proceed on a certain initial state and thus neglect the sum over target states) but it is useful in two respects. Firstly, it simplifies the computation of the rate and therefore is utilized in all \textit{astrophysical} compound reaction codes. Secondly, it allows us to easily find a reciprocity relation between forward and inverse rate by remembering that $E\sigma^\mathrm{eff}$ obeys reciprocity between forward and inverse reaction due to detailed balance. It should be noted that only \textit{stellar} reactivities $\mathcal{R}^*$ (and thus stellar rates $r^*)$ obey reciprocity (as long as detailed balance is applicable) whereas rates derived from ground state cross sections $\sigma^\mathrm{lab}=\sum_j \sigma_{0j}$ do not, unless the SEF is equal to unity in the given direction.

For reactions $1+2\rightarrow 3+4$ with target nucleus 1, projectile 2, final nucleus 3, and ejectile 4, the relation between backward and forward stellar reactivity is given by \cite{ilibook,adndt}
\begin{equation}
\label{eq:revrate}
\mathcal{R}^{*}_{34}=\frac{(2J^0_{2}+1)}{(2J^0_{4}+1)}\left( 
\frac{A_{1}A_{2}}{A_{3}A_{4}}\right) 
^{3/2}\frac{G_1}{G_{3}}e^{-\frac{Q_{12}}{kT}} \mathcal{R}^{*}_{12}\quad ,
\end{equation}
where $Q_{12}$ is the reaction $Q$ value, $J^0$ denote ground state spins, and $G$ are nuclear partition functions summing over states $i$ and integrating over a level density $\rho$ above
the last discrete state $m$ included \cite{hol76,adndt}:
\begin{eqnarray}
G(T)= &  & \sum _{i=0}^{m}(2J_{i}+1)
e^{-\frac{E_{i}}{kT}} \label{eq:partfunc}\\
 &  & +\int\limits _{E_m}^{E^\mathrm{max}} \sum _{J,
\pi}(2J+1)
e^{-\frac{E}{kT}}\rho \left( E,J,\pi \right) \,dE 
\quad .\nonumber 
\end{eqnarray}
This partition function also appears in Eq.\ \ref{eq:simprate} where it is sufficient to compute it once and separately from the rate integration.

Stellar capture reactions $1+2\rightarrow 3+\gamma$ are related to stellar photodisintegration by \cite{hol76,adndt}
\begin{equation}
\label{eq:invphot}
\mathcal{R}^*_{3\gamma }=(2J^0_{2}+1)\left( \frac{A_{1}A_{2}}{A_{3}}\right) 
^{3/2}\left( \frac{kT}{2\pi \hbar ^2}\right)^{3/2}\frac{G_{1}}
{G_{3}} e^{-\frac{Q_{12}}{kT}} \mathcal{R}^{*}_{12}\quad .
\end{equation}

\subsection{Reaction $Q$ value and stellar enhancement factor}
\label{sec:sef}

Figure \ref{fig:eff} shows a sketch of the energetically allowed transitions included in the effective cross section defined by Eq.\ (\ref{eq:effcs}). It is obvious that there are more transitions possible to and from states of the nucleus being the final nucleus in a reaction with positive $Q$ value. Therefore, assuming a similar level structure in all involved nuclei, it is expected that the SEF (see Eq.\ \ref{eq:sef}) of a given reaction will be smaller for the exothermic direction $f_\mathrm{forw}$ than for the endothermic one $f_\mathrm{rev}$ (here we define the forward reaction to be the one with positive $Q$ value and the reverse reaction having negative $Q$ value):
\begin{equation}
f_\mathrm{rev}>f_\mathrm{forw} \quad.
\end{equation}
This is especially pronounced in photodisintegration reactions due to the many possible $\gamma$ transitions \cite{UtsZil,moh07}.
In consequence, aiming at performing a measurement as close as possible to the stellar value, an exothermic reaction should be chosen.

Another impact of the $Q$ value is found by inspection of Eqs.\ (\ref{eq:revrate}) and (\ref{eq:invphot}) where the $Q$ value appears in an exponential. For numerical consistency and to obtain proper equilibrium abundances when forward and reverse reaction are both fast and in equilibrium, reaction network codes avoid employing separate rates for the two directions but rather make use of these equations. Taking an endothermic reaction as starting point for application of the equations would lead to a large value of the exponential term, amplifying any numerical errors inherent in the original rate and in the $Q$ value. This is mainly important when dealing with rate fits. In many astrophysical reaction network codes, the rates are implemented not as large tables but as fits with a smaller number of parameters per reaction. Any deficiency in the fit would be amplified when computing an exothermic rate from an endothermic one.

For the above reasons, it was commonly assumed that it is always preferable to determine the cross section and rate of an exothermic reaction and not those of an endothermic one. Here, we want to correct that notion by showing that
there are cases for which
\begin{equation}
f_\mathrm{rev}<f_\mathrm{forw} \quad.
\end{equation}
The basic idea is to realize that although there are more transitions energetically possible to the final states of exothermic reactions, some of them may be suppressed and thus not contributing. Of course, it is obvious that not all transitions are of equal strength. Quantum mechanical spin and parity selection rules and centrifugal barriers (or lack thereof) may prefer certain transitions over others. This will be important in reactions with small $\left| Q \right|$ and in nuclei with large level spacings. In both cases, only a small number of transitions will be possible and the spins can give larger weight to an even smaller subset. However, for reactions with sizeable $Q$ values or involving nuclei with high level densities this suppression due to spins will not be sufficient because there will always be a number of states with matching spins.

Transitions to higher lying excited states of a nucleus proceed at lower relative energy. Except for s-wave neutrons, also transitions at lower relative energy will be weaker than those at larger relative energy. If the suppression of transitions with smaller relative energy is different in the entrance and exit channel of the reaction, this may also result in $f_\mathrm{rev}<f_\mathrm{forw}$. The strongest suppression for charged particles is due to the Coulomb barrier. Having different Coulomb barriers in the entrance and exit channel, e.g.\, in (n,p) or (p,$\alpha$) reactions, can more strongly suppress the transitions to the nucleus with higher Coulomb barrier than to the one with lower Coulomb barrier. With respect to Fig.\ \ref{fig:eff} and assuming, e.g., a reaction A(n,$\alpha$)F this means that most transitions to states in nucleus F are suppressed and the contributing transitions may be fewer than those accessing states in nucleus A.
 
This Coulomb suppression of the SEF is a general principle almost independent of nuclear structure and will act for a large range of nuclei. Whether the suppression is strong enough to yield $f_\mathrm{rev}<f_\mathrm{forw}$ depends on the size of the $Q$ value relative to the Coulomb barrier, i.e. the effect can occur in a reaction provided that
there are different Coulomb barriers in the entrance and exit channel and $\left| Q\right|$ is low compared to the Coulomb barrier. The strongest impact is to be expected when the forward reaction involves neutrons in the entrance channel which form a compound state by s-waves on excited target states and charged particles experiencing a high Coulomb barrier in the exit channel. As discussed in Sec.\ \ref{sec:astrorate} the reaction $^{85}$Rb(p,n)$^{85}$Sr is an excellent example for such a case.
A quantitative exploration of the suppression effect across the nuclear chart is given in the following section.

\begin{figure}
\resizebox{\columnwidth}{!}{\rotatebox{270}{\includegraphics[clip=]{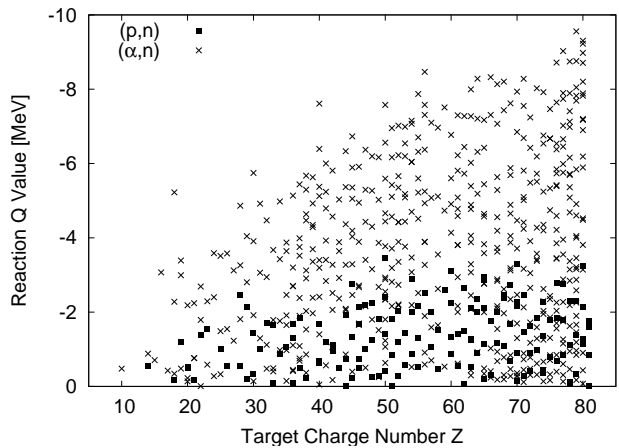}}}
\caption{Reaction $Q$ values for (p,n) and ($\alpha$,n) reactions with $f_\mathrm{rev}<f_\mathrm{forw}$.\label{fig:qdep}
}
\end{figure}

Not only theoretically interesting, this Coulomb suppression effect is also important for experiments
because it allows to directly determine an astrophysically relevant rate by measuring in the
direction of suppressed SEF. The above mentioned complication of fitting rates
with negative $Q$ values can be circumvented by directly applying detailed balance and numerically
computing the rate for the forward reaction before performing a fit. This is possible when $f_\mathrm{rev}\approx 1$. Subsequently, fits for both rates can be obtained in the standard way. As an example, an application of this procedure to the rate of $^{85}$Rb(p,n)$^{85}$Sr is shown in Sec.\ \ref{sec:astrorate}.

\subsection{Exploration of the SEF suppression across the nuclear chart}
\label{sec:theoryresults}

In this section, we quantitatively study the SEF suppression introduced and discussed in the previous section. Using NON-SMOKER results \cite{nonsmoker,adndt} we compared $f_\mathrm{forw}$ and $f_\mathrm{rev}$ for reactions involving light projectiles (nucleons, $\alpha$) and targets from Ne to Bi between the proton and neutron driplines. To avoid trivial cases, only reactions with $f_\mathrm{forw}/f_\mathrm{rev} \geq 1.1$ are considered. Furthermore, the SEF were computed for $T\leq 4.5$ GK to find cases important in most nucleosynthesis environments and to eliminate cases only occurring at high temperature. Because of our aim to provide guidance for experiments, we further only focus on examples with $f_\mathrm{rev} \leq 1.5$. Even with these restrictions we find 1200 reactions exhibiting such a strong suppression effect that $f_\mathrm{rev}<f_\mathrm{forw}$.

To check the dependence on the Coulomb barrier, Fig.\ \ref{fig:qdep}
shows the obtained range of $Q$ values still yielding $f_\mathrm{rev}<f_\mathrm{forw}$ as a function of target charge $Z$ for (p,n) and ($\alpha$,n) reactions with
negative $Q$ values. It can be clearly seen that
larger maximal $\left| Q\right|$ is allowed with increasing charge $Z$. The different increase in permitted maximal $\left| Q\right|$ is different for the two reactions, reflecting the difference in the height of the acting Coulomb barriers. Below each maximally allowed $\left| Q\right|$ for each given charge, there is a range of other values. This scatter is mainly caused by the available $Q$ values (as defined by the masses of the nuclei) and not by other effects such as spins and parities of the involved nuclei. Although the strengths of the involved transitions also depend on spin and parity of the initial and final state, Coulomb repulsion dominates the suppression when the interaction energy is small, as it is the case for astrophysically relevant energies.

\begin{figure}
\resizebox{\columnwidth}{!}{\rotatebox{270}{\includegraphics[clip=]{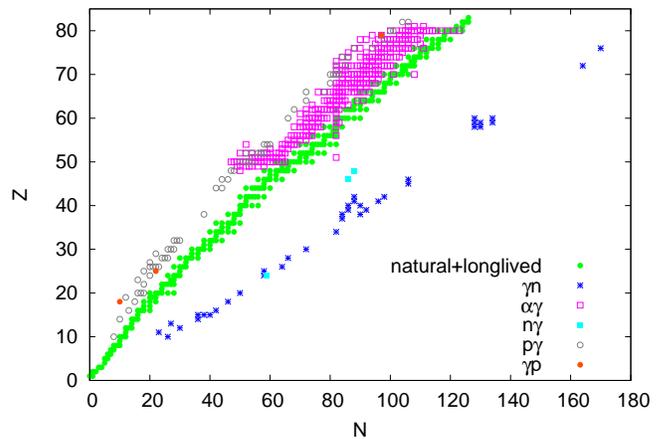}}}
\caption{(Color online) Targets for endothermic reactions with $f_\mathrm{rev}<f_\mathrm{forw}$ in the nuclear chart, where charge is denoted by $Z$ and neutron number by $N$. The reaction type is given by the label. Only capture or photodisintegration reactions are shown. Also printed for orientation are stable and longlived nuclides.\label{fig:gamma}
}
\end{figure}
\begin{figure}
\resizebox{\columnwidth}{!}{\rotatebox{270}{\includegraphics[clip=]{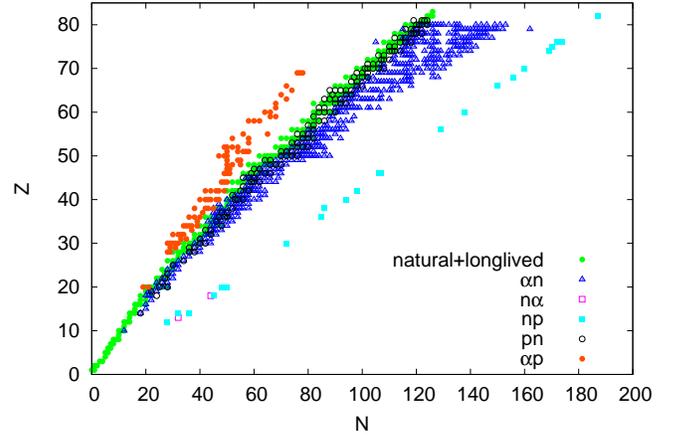}}}
\caption{(Color online) Targets for endothermic reactions with $f_\mathrm{rev}<f_\mathrm{forw}$ in the nuclear chart, where charge is denoted by $Z$ and neutron number by $N$. The reaction type is given by the label. Only reactions without $\gamma$ channels are shown. Also printed for orientation are stable and longlived nuclides.\label{fig:particle}
}
\end{figure}

\begin{table}
\caption{\label{tab:ag} Targets for ($\alpha$,$\gamma$) reactions with negative $Q$ value but smaller SEF than their inverse reaction. Stable or longlived targets are in italics.}
\begin{ruledtabular}
\begin{tabular}{cccccccc}
$^{98}$Cd & $^{104}$Te & $^{130}$Ce & $^{143}$Eu & $^{150}$Er & $^{153}$Lu & $^{173}$W & $^{178}$Pt \\
$^{98}$In & $^{106}$Te & $^{131}$Ce & $^{145}$Eu & $^{151}$Er & $^{155}$Lu & $^{174}$W & $^{179}$Pt \\
$^{99}$In & $^{109}$Te & $^{132}$Ce & $^{147}$Eu & $^{152}$Er & $^{156}$Lu & $^{176}$W & $^{180}$Pt \\
$^{101}$In & $^{110}$Te & $^{133}$Ce & $^{149}$Eu & $^{153}$Er & $^{162}$Lu & $^{177}$W & $^{181}$Pt \\
$^{102}$In & $^{111}$Te & $^{134}$Ce & $^{142}$Gd & $^{154}$Er & $^{163}$Lu & $^{161}$Re & $^{182}$Pt \\
$^{103}$In & $^{112}$Te & $^{135}$Ce & $^{143}$Gd & $^{155}$Er & $^{166}$Lu & $^{167}$Re & $^{183}$Pt \\
$^{105}$In & $^{114}$Te & \textit{$^\textit{140}$Ce} & $^{144}$Gd & $^{156}$Er & $^{167}$Lu & $^{171}$Re & $^{184}$Pt \\
$^{107}$In & $^{115}$Te & $^{129}$Pr & $^{145}$Gd & $^{157}$Er & $^{154}$Hf & $^{172}$Re & $^{185}$Pt \\
$^{97}$Sn & $^{116}$Te & $^{131}$Pr & $^{146}$Gd & $^{158}$Er & $^{156}$Hf & $^{175}$Re & $^{186}$Pt \\
$^{98}$Sn & $^{117}$Te & $^{133}$Pr & $^{147}$Gd & $^{159}$Er & $^{157}$Hf & $^{179}$Re & $^{187}$Pt \\
$^{99}$Sn & $^{113}$I & $^{135}$Pr & $^{148}$Gd & $^{160}$Er & $^{158}$Hf & $^{164}$Os & $^{173}$Au \\
$^{100}$Sn & $^{117}$I & \textit{$^\textit{141}$Pr} & $^{149}$Gd & $^{161}$Er & $^{159}$Hf & $^{166}$Os & $^{176}$Au \\
$^{101}$Sn & $^{119}$I & $^{142}$Pr & \textit{$^\textit{153}$Gd} & \textit{$^\textit{162}$Er} & $^{161}$Hf & $^{170}$Os & $^{182}$Au \\
$^{102}$Sn & $^{106}$Xe & $^{132}$Nd & \textit{$^\textit{155}$Gd} & \textit{$^\textit{163}$Er} & $^{162}$Hf & $^{172}$Os & $^{184}$Au \\
$^{103}$Sn & $^{118}$Xe & $^{133}$Nd & $^{147}$Tb & $^{153}$Tm & $^{163}$Hf & $^{173}$Os & $^{185}$Au \\
$^{104}$Sn & $^{119}$Xe & $^{134}$Nd & $^{148}$Tb & $^{155}$Tm & $^{164}$Hf & $^{174}$Os & $^{189}$Au \\
$^{105}$Sn & $^{120}$Xe & $^{135}$Nd & $^{149}$Tb & $^{158}$Tm & $^{165}$Hf & $^{175}$Os & $^{180}$Hg \\
$^{106}$Sn & $^{121}$Xe & $^{136}$Nd & $^{142}$Dy & $^{159}$Tm & $^{166}$Hf & $^{176}$Os & $^{182}$Hg \\
$^{107}$Sn & $^{122}$Xe & $^{137}$Nd & $^{144}$Dy & $^{161}$Tm & $^{167}$Hf & $^{177}$Os & $^{184}$Hg \\
$^{108}$Sn & $^{123}$Xe & $^{138}$Nd & $^{146}$Dy & $^{163}$Tm & $^{169}$Hf & $^{178}$Os & $^{185}$Hg \\
$^{109}$Sn & $^{121}$Cs & $^{139}$Nd & $^{147}$Dy & $^{165}$Tm & $^{170}$Hf & $^{179}$Os & $^{186}$Hg \\
$^{110}$Sn & $^{123}$Cs & \textit{$^\textit{142}$Nd} & $^{148}$Dy & $^{152}$Yb & $^{171}$Hf & $^{180}$Os & $^{187}$Hg \\
$^{111}$Sn & $^{125}$Cs & $^{131}$Pm & $^{149}$Dy & $^{153}$Yb & $^{172}$Hf & $^{181}$Os & $^{188}$Hg \\
\textit{$^\textit{112}$Sn} & $^{127}$Cs & $^{133}$Pm & $^{150}$Dy & $^{154}$Yb & $^{173}$Hf & $^{182}$Os & $^{189}$Hg \\
\textit{$^\textit{113}$Sn} & $^{123}$Ba & $^{137}$Pm & $^{151}$Dy & $^{156}$Yb & $^{155}$Ta & $^{183}$Os & $^{190}$Hg \\
\textit{$^\textit{114}$Sn} & $^{124}$Ba & \textit{$^\textit{143}$Pm} & $^{152}$Dy & $^{157}$Yb & $^{165}$Ta & \textit{$^\textit{187}$Os} & $^{191}$Hg \\
\textit{$^\textit{115}$Sn} & $^{125}$Ba & $^{134}$Sm & $^{154}$Dy & $^{158}$Yb & $^{167}$Ta & $^{172}$Ir & $^{192}$Hg \\
$^{105}$Sb & $^{126}$Ba & $^{136}$Sm & $^{155}$Dy & $^{159}$Yb & $^{169}$Ta & $^{175}$Ir & $^{193}$Hg \\
$^{106}$Sb & $^{127}$Ba & $^{137}$Sm & \textit{$^\textit{156}$Dy} & $^{160}$Yb & $^{158}$W & $^{177}$Ir & $^{194}$Hg \\
$^{107}$Sb & $^{128}$Ba & $^{138}$Sm & \textit{$^\textit{157}$Dy} & $^{161}$Yb & $^{160}$W & $^{179}$Ir & $^{195}$Hg \\
$^{108}$Sb & $^{129}$Ba & $^{139}$Sm & \textit{$^\textit{159}$Dy} & $^{162}$Yb & $^{164}$W & $^{181}$Ir & \textit{$^\textit{197}$Hg} \\
$^{109}$Sb & \textit{$^\textit{130}$Ba} & $^{140}$Sm & $^{149}$Ho & $^{163}$Yb & $^{166}$W & $^{183}$Ir & \textit{$^\textit{199}$Hg} \\
$^{111}$Sb & \textit{$^\textit{138}$Ba} & $^{141}$Sm & $^{150}$Ho & $^{164}$Yb & $^{167}$W & $^{187}$Ir & \textit{$^\textit{201}$Hg} \\
$^{112}$Sb & $^{129}$La & $^{142}$Sm & $^{151}$Ho & $^{165}$Yb & $^{168}$W & $^{168}$Pt & \textit{$^\textit{203}$Hg} \\
$^{113}$Sb & $^{131}$La & $^{143}$Sm & $^{154}$Ho & $^{166}$Yb & $^{169}$W & $^{170}$Pt & $^{187}$Tl \\
$^{115}$Sb & $^{126}$Ce & \textit{$^\textit{144}$Sm} & $^{155}$Ho & $^{167}$Yb & $^{170}$W & $^{174}$Pt & $^{189}$Tl \\
$^{133}$Sb & $^{128}$Ce & \textit{$^\textit{145}$Sm} & $^{157}$Ho & \textit{$^\textit{171}$Yb} & $^{171}$W & $^{176}$Pt & $^{192}$Tl \\
$^{102}$Te & $^{129}$Ce & $^{137}$Eu & $^{159}$Ho & $^{178}$Yb & $^{172}$W & $^{177}$Pt & 
\end{tabular}
\end{ruledtabular}
\end{table}

\begin{table}
\caption{\label{tab:pg} Targets for (p,$\gamma$) reactions with negative $Q$ value but smaller SEF than their inverse reaction. No stable or longlived targets were found.}
\begin{ruledtabular}
\begin{tabular}{cccccccc}
$^{18}$Ne & $^{39}$V & $^{53}$Ni & $^{76}$Sr & $^{103}$Sn & $^{114}$Xe & $^{152}$Yb & $^{180}$Hg \\
$^{24}$Si & $^{44}$Cr & $^{54}$Ni & $^{86}$Ru & $^{104}$Sn & $^{126}$Nd & $^{153}$Yb & $^{181}$Hg \\
$^{29}$S & $^{43}$Mn & $^{51}$Cu & $^{88}$Ru & $^{104}$Te & $^{130}$Sm & $^{154}$Hf & $^{186}$Pb \\
$^{33}$Ar & $^{46}$Fe & $^{56}$Zn & $^{90}$Pd & $^{106}$Te & $^{136}$Gd & $^{157}$Hf & $^{188}$Pb \\
$^{31}$K & $^{47}$Fe & $^{57}$Zn & $^{92}$Pd & $^{108}$Te & $^{138}$Dy & $^{158}$W &  \\
$^{36}$Ca & $^{48}$Fe & $^{58}$Zn & $^{95}$Cd & $^{109}$Te & $^{144}$Dy & $^{160}$W &  \\
$^{37}$Ca & $^{49}$Fe & $^{60}$Ge & $^{96}$Cd & $^{110}$Te & $^{148}$Er & $^{164}$Os &  \\
$^{38}$Ca & $^{47}$Co & $^{61}$Ge & $^{100}$Sn & $^{112}$Xe & $^{150}$Yb & $^{170}$Os &  \\
$^{40}$Ti & $^{52}$Ni & $^{62}$Ge & $^{102}$Sn & $^{113}$Xe & $^{151}$Yb & $^{176}$Pt & 
\end{tabular}
\end{ruledtabular}
\end{table}

\begin{table}
\caption{\label{tab:gn} Targets for ($\gamma$,n) reactions with negative $Q$ value but smaller SEF than their inverse reaction. No stable or longlived targets were found.}
\begin{ruledtabular}
\begin{tabular}{cccccccc}
$^{36}$Ne & $^{51}$P & $^{70}$Ca & $^{102}$Zn & $^{125}$Y & $^{137}$Nb & $^{186}$Ce & $^{188}$Nd \\
$^{34}$Na & $^{53}$P & $^{82}$Cr & $^{116}$Se & $^{131}$Y & $^{130}$Mo & $^{188}$Ce & $^{194}$Nd \\
$^{42}$Mg & $^{55}$P & $^{83}$Mn & $^{121}$Rb & $^{126}$Zr & $^{140}$Mo & $^{187}$Pr & $^{236}$Hf \\
$^{40}$Al & $^{58}$S & $^{90}$Fe & $^{122}$Sr & $^{130}$Zr & $^{151}$Rh & $^{189}$Pr & $^{246}$Os \\
$^{50}$Si & $^{64}$Ar & $^{94}$Ni & $^{128}$Sr & $^{129}$Nb & $^{152}$Pd & $^{193}$Pr & $^{176}$Au
\end{tabular}
\end{ruledtabular}
\end{table}

\begin{table}
\caption{\label{tab:gpnang} Targets for ($\gamma$,p), (n,$\alpha$), and (n,$\gamma$) reactions with negative $Q$ value but smaller SEF than their inverse reaction. No stable or longlived targets were found.}
\begin{ruledtabular}
\begin{tabular}{cccccccc}
($\gamma$,p): &&&&&&&\\
&$^{28}$Ar & $^{47}$Mn & $^{176}$Au &    &  &  & \\
(n,$\alpha$): &&&&&&&\\
&$^{45}$Al & $^{62}$Ar &  &  &  &  &  \\
(n,$\gamma$): &&&&&&&\\
&$^{83}$Cr & $^{132}$Pd & $^{136}$Cd &  &  &  &  
\end{tabular}
\end{ruledtabular}
\end{table}

\begin{table}
\caption{\label{tab:pn} Targets for (p,n) reactions with negative $Q$ value but smaller SEF than their inverse reaction. Stable or longlived targets are in italics.}
\begin{ruledtabular}
\begin{tabular}{cccccccc}
$^{32}$Si & $^{77}$As & $^{101}$Rh & \textit{$^\textit{123}$Sb} & \textit{$^\textit{144}$Nd} & \textit{$^\textit{159}$Tb} & $^{177}$Lu & \textit{$^\textit{197}$Pt} \\
$^{42}$Ar & \textit{$^\textit{82}$Se} & \textit{$^\textit{103}$Rh} & $^{125}$Sb & \textit{$^\textit{146}$Nd} & $^{161}$Tb & \textit{$^\textit{179}$Hf} & \textit{$^\textit{198}$Pt} \\
$^{41}$K & \textit{$^\textit{81}$Br} & $^{105}$Rh & \textit{$^\textit{128}$Te} & \textit{$^\textit{148}$Nd} & \textit{$^\textit{164}$Dy} & \textit{$^\textit{180}$Hf} & $^{200}$Pt \\
\textit{$^\textit{45}$Ca} & \textit{$^\textit{83}$Kr} & \textit{$^\textit{105}$Pd} & \textit{$^\textit{130}$Te} & \textit{$^\textit{150}$Nd} & $^{166}$Dy & $^{182}$Hf & $^{195}$Au \\
\textit{$^\textit{48}$Ca} & \textit{$^\textit{85}$Kr} & \textit{$^\textit{107}$Pd} & $^{132}$Te & \textit{$^\textit{145}$Pm} & $^{163}$Ho & $^{179}$Ta & \textit{$^\textit{197}$Au} \\
$^{47}$Sc & \textit{$^\textit{86}$Kr} & \textit{$^\textit{110}$Pd} & $^{127}$I & \textit{$^\textit{147}$Pm} & \textit{$^\textit{165}$Ho} & \textit{$^\textit{181}$Ta} & $^{199}$Au \\
\textit{$^\textit{49}$Ti} & \textit{$^\textit{85}$Rb} & $^{112}$Pd & $^{129}$I & \textit{$^\textit{152}$Sm} & \textit{$^\textit{165}$Er} & $^{184}$W & \textit{$^\textit{200}$Hg} \\
$^{51}$V & $^{87}$Rb & \textit{$^\textit{107}$Ag} & \textit{$^\textit{132}$Xe} & \textit{$^\textit{154}$Sm} & \textit{$^\textit{168}$Er} & $^{185}$W & \textit{$^\textit{201}$Hg} \\
\textit{$^\textit{55}$Mn} & $^{90}$Sr & \textit{$^\textit{109}$Ag} & \textit{$^\textit{134}$Xe} & $^{156}$Sm & \textit{$^\textit{170}$Er} & $^{186}$W & \textit{$^\textit{202}$Hg} \\
$^{60}$Fe & \textit{$^\textit{93}$Zr} & \textit{$^\textit{114}$Cd} & \textit{$^\textit{136}$Xe} & $^{149}$Eu & $^{167}$Tm & $^{188}$W & \textit{$^\textit{204}$Hg} \\
\textit{$^\textit{64}$Ni} & \textit{$^\textit{94}$Zr} & \textit{$^\textit{116}$Cd} & $^{131}$Cs & \textit{$^\textit{151}$Eu} & \textit{$^\textit{169}$Tm} & \textit{$^\textit{185}$Re} & $^{200}$Tl \\
$^{66}$Ni & \textit{$^\textit{96}$Zr} & $^{118}$Cd & \textit{$^\textit{133}$Cs} & \textit{$^\textit{153}$Eu} & $^{171}$Tm & \textit{$^\textit{187}$Re} & \textit{$^\textit{203}$Tl} \\
\textit{$^\textit{65}$Cu} & \textit{$^\textit{93}$Nb} & \textit{$^\textit{113}$In} & $^{135}$Cs & $^{155}$Eu & \textit{$^\textit{171}$Yb} & \textit{$^\textit{190}$Os} & \textit{$^\textit{204}$Tl} \\
$^{67}$Cu & \textit{$^\textit{97}$Mo} & \textit{$^\textit{115}$In} & \textit{$^\textit{138}$Ba} & \textit{$^\textit{153}$Gd} & \textit{$^\textit{172}$Yb} & \textit{$^\textit{192}$Os} & \textit{$^\textit{205}$Tl} \\
\textit{$^\textit{70}$Zn} & \textit{$^\textit{100}$Mo} & \textit{$^\textit{120}$Sn} & $^{137}$La & \textit{$^\textit{158}$Gd} & \textit{$^\textit{174}$Yb} & $^{194}$Os &  \\
$^{72}$Zn & \textit{$^\textit{99}$Tc} & \textit{$^\textit{122}$Sn} & \textit{$^\textit{139}$La} & \textit{$^\textit{160}$Gd} & \textit{$^\textit{176}$Yb} & $^{189}$Ir &  \\
\textit{$^\textit{71}$Ga} & \textit{$^\textit{103}$Ru} & \textit{$^\textit{124}$Sn} & \textit{$^\textit{142}$Ce} & $^{153}$Tb & $^{178}$Yb & \textit{$^\textit{191}$Ir} &  \\
\textit{$^\textit{76}$Ge} & \textit{$^\textit{104}$Ru} & $^{126}$Sn & $^{144}$Ce & $^{155}$Tb & $^{173}$Lu & \textit{$^\textit{193}$Ir} &  \\
\textit{$^\textit{75}$As} & $^{106}$Ru & \textit{$^\textit{121}$Sb} & \textit{$^\textit{141}$Pr} & $^{157}$Tb & \textit{$^\textit{175}$Lu} & \textit{$^\textit{196}$Pt} & 
\end{tabular}
\end{ruledtabular}
\end{table}

\begin{table}
\caption{\label{tab:np} Targets for (n,p) reactions with negative $Q$ value but smaller SEF than their inverse reaction. No stable or longlived targets were found.}
\begin{ruledtabular}
\begin{tabular}{cccccccc}
$^{40}$Mg & $^{63}$Ar & $^{102}$Zn & $^{134}$Zr & $^{153}$Pd & $^{216}$Dy & $^{243}$W & $^{250}$Os \\
$^{46}$Si & $^{68}$Ca & $^{121}$Kr & $^{140}$Mo & $^{185}$Ba & $^{224}$Er & $^{245}$Re & $^{269}$Pb \\
$^{50}$Si & $^{70}$Ca & $^{124}$Sr & $^{152}$Pd & $^{198}$Nd & $^{230}$Yb & $^{248}$Os & 
\end{tabular}
\end{ruledtabular}
\end{table}

\begingroup
\squeezetable
\begin{table}
\caption{\label{tab:an} Targets for ($\alpha$,n) reactions with negative $Q$ value but smaller SEF than their inverse reaction. Stable or longlived targets are in italics.}
\begin{ruledtabular}
\begin{tabular}{cccccccc}
\textit{$^\textit{22}$Ne} & \textit{$^\textit{83}$Kr} & $^{106}$Ru & $^{133}$Te & $^{160}$Pm & $^{175}$Tm & $^{190}$W & \textit{$^\textit{198}$Pt} \\
$^{32}$Si & \textit{$^\textit{84}$Kr} & $^{108}$Ru & $^{134}$Te & $^{163}$Pm & $^{176}$Tm & $^{192}$W & $^{199}$Pt \\
$^{35}$P & \textit{$^\textit{86}$Kr} & $^{110}$Ru & $^{135}$Te & $^{164}$Pm & $^{178}$Tm & $^{197}$W & $^{200}$Pt \\
$^{36}$S & $^{87}$Kr & $^{112}$Ru & $^{136}$Te & $^{167}$Pm & $^{179}$Tm & $^{200}$W & $^{201}$Pt \\
$^{39}$Cl & $^{88}$Kr & $^{102}$Rh & $^{137}$Te & \textit{$^\textit{154}$Sm} & $^{180}$Tm & $^{201}$W & $^{202}$Pt \\
\textit{$^\textit{38}$Ar} & $^{90}$Kr & \textit{$^\textit{103}$Rh} & $^{138}$Te & $^{156}$Sm & $^{181}$Tm & $^{202}$W & $^{204}$Pt \\
\textit{$^\textit{39}$Ar} & $^{82}$Rb & $^{104}$Rh & $^{129}$I & $^{158}$Sm & $^{182}$Tm & $^{203}$W & $^{205}$Pt \\
\textit{$^\textit{40}$Ar} & $^{83}$Rb & $^{105}$Rh & $^{131}$I & $^{155}$Eu & $^{183}$Tm & $^{206}$W & $^{206}$Pt \\
$^{40}$K & $^{84}$Rb & $^{106}$Rh & $^{132}$I & $^{157}$Eu & $^{184}$Tm & $^{207}$W & $^{210}$Pt \\
$^{41}$K & \textit{$^\textit{85}$Rb} & $^{107}$Rh & $^{133}$I & $^{158}$Eu & $^{185}$Tm & $^{210}$W & $^{212}$Pt \\
$^{43}$K & $^{86}$Rb & $^{109}$Rh & $^{135}$I & $^{159}$Eu & $^{186}$Tm & $^{212}$W & $^{213}$Pt \\
\textit{$^\textit{44}$Ca} & $^{87}$Rb & $^{111}$Rh & $^{136}$I & $^{160}$Eu & $^{188}$Tm & $^{214}$W & $^{218}$Pt \\
\textit{$^\textit{46}$Ca} & $^{88}$Rb & $^{114}$Rh & $^{137}$I & $^{162}$Eu & $^{191}$Tm & $^{190}$Re & $^{220}$Pt \\
\textit{$^\textit{48}$Ca} & $^{90}$Rb & \textit{$^\textit{108}$Pd} & \textit{$^\textit{132}$Xe} & $^{163}$Eu & $^{195}$Tm & $^{191}$Re & $^{221}$Pt \\
\textit{$^\textit{45}$Sc} & $^{91}$Rb & \textit{$^\textit{109}$Pd} & \textit{$^\textit{134}$Xe} & $^{164}$Eu & $^{177}$Yb & $^{192}$Re & $^{222}$Pt \\
$^{47}$Sc & $^{83}$Sr & \textit{$^\textit{110}$Pd} & \textit{$^\textit{136}$Xe} & $^{165}$Eu & $^{178}$Yb & $^{194}$Re & $^{223}$Pt \\
$^{49}$Sc & \textit{$^\textit{85}$Sr} & $^{111}$Pd & $^{138}$Xe & $^{166}$Eu & $^{186}$Yb & $^{200}$Re & $^{224}$Pt \\
\textit{$^\textit{48}$Ti} & \textit{$^\textit{86}$Sr} & $^{112}$Pd & $^{139}$Xe & $^{167}$Eu & $^{188}$Yb & $^{201}$Re & $^{194}$Au \\
\textit{$^\textit{50}$Ti} & \textit{$^\textit{88}$Sr} & $^{114}$Pd & $^{140}$Xe & $^{169}$Eu & $^{190}$Yb & $^{209}$Re & $^{198}$Au \\
$^{52}$Ti & $^{90}$Sr & $^{116}$Pd & $^{141}$Xe & $^{170}$Eu & $^{191}$Yb & $^{210}$Re & $^{199}$Au \\
$^{51}$V & $^{91}$Sr & \textit{$^\textit{109}$Ag} & $^{142}$Xe & $^{173}$Eu & $^{194}$Yb & $^{211}$Re & $^{201}$Au \\
$^{53}$V & $^{92}$Sr & $^{111}$Ag & $^{146}$Xe & \textit{$^\textit{158}$Gd} & $^{196}$Yb & $^{212}$Re & $^{205}$Au \\
\textit{$^\textit{52}$Cr} & $^{93}$Sr & $^{113}$Ag & $^{135}$Cs & \textit{$^\textit{160}$Gd} & $^{198}$Yb & $^{213}$Re & $^{213}$Au \\
\textit{$^\textit{54}$Cr} & $^{94}$Sr & $^{115}$Ag & $^{137}$Cs & $^{161}$Gd & $^{200}$Yb & $^{215}$Re & $^{214}$Au \\
$^{56}$Cr & $^{96}$Sr & $^{116}$Ag & $^{138}$Cs & $^{162}$Gd & $^{178}$Lu & $^{217}$Re & $^{215}$Au \\
\textit{$^\textit{55}$Mn} & $^{86}$Y & $^{117}$Ag & $^{139}$Cs & $^{171}$Gd & $^{179}$Lu & $^{181}$Os & $^{216}$Au \\
$^{56}$Mn & $^{90}$Y & \textit{$^\textit{112}$Cd} & $^{140}$Cs & $^{162}$Tb & $^{183}$Lu & \textit{$^\textit{192}$Os} & $^{217}$Au \\
$^{57}$Mn & $^{91}$Y & \textit{$^\textit{114}$Cd} & $^{141}$Cs & $^{163}$Tb & $^{184}$Lu & $^{193}$Os & $^{218}$Au \\
\textit{$^\textit{58}$Fe} & $^{92}$Y & \textit{$^\textit{116}$Cd} & $^{145}$Cs & $^{164}$Tb & $^{185}$Lu & $^{194}$Os & $^{219}$Au \\
$^{60}$Fe & $^{93}$Y & $^{118}$Cd & \textit{$^\textit{136}$Ba} & $^{166}$Tb & $^{186}$Lu & $^{195}$Os & $^{220}$Au \\
$^{61}$Co & $^{94}$Y & $^{120}$Cd & \textit{$^\textit{138}$Ba} & $^{168}$Tb & $^{187}$Lu & $^{196}$Os & $^{221}$Au \\
$^{63}$Co & $^{95}$Y & $^{117}$In & $^{140}$Ba & $^{169}$Tb & $^{188}$Lu & $^{198}$Os & $^{222}$Au \\
\textit{$^\textit{64}$Ni} & $^{97}$Y & $^{119}$In & $^{141}$Ba & $^{170}$Tb & $^{189}$Lu & $^{199}$Os & $^{223}$Au \\
$^{66}$Ni & \textit{$^\textit{90}$Zr} & $^{121}$In & $^{142}$Ba & $^{171}$Tb & $^{194}$Lu & $^{200}$Os & $^{224}$Au \\
$^{68}$Ni & \textit{$^\textit{92}$Zr} & $^{123}$In & $^{143}$Ba & $^{173}$Tb & $^{195}$Lu & $^{202}$Os & $^{225}$Au \\
$^{67}$Cu & \textit{$^\textit{94}$Zr} & $^{125}$In & $^{144}$Ba & $^{175}$Tb & $^{197}$Lu & $^{203}$Os & $^{226}$Au \\
$^{69}$Cu & \textit{$^\textit{95}$Zr} & $^{127}$In & $^{145}$Ba & $^{176}$Tb & $^{198}$Lu & $^{204}$Os & $^{227}$Au \\
\textit{$^\textit{68}$Zn} & \textit{$^\textit{96}$Zr} & \textit{$^\textit{120}$Sn} & $^{146}$Ba & $^{179}$Tb & $^{201}$Lu & $^{208}$Os & $^{228}$Au \\
\textit{$^\textit{70}$Zn} & $^{97}$Zr & \textit{$^\textit{122}$Sn} & $^{147}$Ba & \textit{$^\textit{164}$Dy} & $^{203}$Lu & $^{209}$Os & $^{229}$Au \\
$^{72}$Zn & $^{98}$Zr & \textit{$^\textit{124}$Sn} & $^{148}$Ba & $^{166}$Dy & $^{183}$Hf & $^{210}$Os & $^{231}$Au \\
$^{74}$Zn & $^{100}$Zr & $^{125}$Sn & $^{150}$Ba & $^{177}$Dy & $^{184}$Hf & $^{211}$Os & $^{241}$Au \\
\textit{$^\textit{71}$Ga} & $^{102}$Zr & $^{126}$Sn & $^{143}$La & $^{179}$Dy & $^{190}$Hf & $^{213}$Os & \textit{$^\textit{201}$Hg} \\
$^{73}$Ga & $^{94}$Nb & $^{127}$Sn & $^{144}$La & $^{184}$Dy & $^{192}$Hf & $^{214}$Os & \textit{$^\textit{202}$Hg} \\
$^{75}$Ga & $^{95}$Nb & $^{128}$Sn & $^{148}$La & $^{167}$Ho & $^{195}$Hf & $^{215}$Os & \textit{$^\textit{204}$Hg} \\
\textit{$^\textit{74}$Ge} & $^{96}$Nb & $^{129}$Sn & $^{150}$La & $^{169}$Ho & $^{196}$Hf & $^{216}$Os & $^{205}$Hg \\
\textit{$^\textit{75}$Ge} & $^{97}$Nb & $^{130}$Sn & $^{146}$Ce & $^{170}$Ho & $^{197}$Hf & \textit{$^\textit{192}$Ir} & $^{206}$Hg \\
\textit{$^\textit{76}$Ge} & $^{98}$Nb & $^{131}$Sn & $^{148}$Ce & $^{172}$Ho & $^{198}$Hf & \textit{$^\textit{193}$Ir} & $^{207}$Hg \\
$^{78}$Ge & $^{99}$Nb & $^{132}$Sn & $^{150}$Ce & $^{173}$Ho & $^{199}$Hf & $^{194}$Ir & $^{208}$Hg \\
$^{77}$As & $^{103}$Nb & $^{133}$Sn & $^{152}$Ce & $^{174}$Ho & $^{200}$Hf & $^{195}$Ir & $^{209}$Hg \\
$^{78}$As & $^{105}$Nb & $^{134}$Sn & $^{154}$Ce & $^{175}$Ho & $^{201}$Hf & $^{196}$Ir & $^{210}$Hg \\
$^{79}$As & \textit{$^\textit{98}$Mo} & $^{136}$Sn & $^{147}$Pr & $^{176}$Ho & $^{204}$Hf & $^{197}$Ir & $^{212}$Hg \\
$^{81}$As & \textit{$^\textit{99}$Mo} & $^{138}$Sn & $^{148}$Pr & $^{177}$Ho & $^{185}$Ta & $^{199}$Ir & $^{214}$Hg \\
\textit{$^\textit{78}$Se} & \textit{$^\textit{100}$Mo} & \textit{$^\textit{123}$Sb} & $^{156}$Pr & $^{178}$Ho & $^{188}$Ta & $^{203}$Ir & $^{215}$Hg \\
\textit{$^\textit{79}$Se} & $^{101}$Mo & $^{125}$Sb & \textit{$^\textit{150}$Nd} & $^{179}$Ho & $^{189}$Ta & $^{211}$Ir & $^{216}$Hg \\
\textit{$^\textit{80}$Se} & $^{102}$Mo & $^{127}$Sb & $^{151}$Nd & $^{181}$Ho & $^{190}$Ta & $^{213}$Ir & $^{222}$Hg \\
\textit{$^\textit{82}$Se} & $^{103}$Mo & $^{128}$Sb & $^{152}$Nd & $^{182}$Ho & $^{191}$Ta & $^{214}$Ir & $^{224}$Hg \\
$^{84}$Se & $^{104}$Mo & $^{129}$Sb & $^{154}$Nd & $^{183}$Ho & $^{192}$Ta & $^{215}$Ir & $^{226}$Hg \\
$^{78}$Br & $^{106}$Mo & $^{130}$Sb & \textit{$^\textit{151}$Pm} & $^{184}$Ho & $^{198}$Ta & $^{216}$Ir & $^{229}$Hg \\
\textit{$^\textit{80}$Br} & \textit{$^\textit{100}$Tc} & $^{131}$Sb & $^{152}$Pm & $^{185}$Ho & $^{199}$Ta & $^{217}$Ir & $^{231}$Hg \\
\textit{$^\textit{81}$Br} & \textit{$^\textit{101}$Tc} & $^{133}$Sb & $^{153}$Pm & \textit{$^\textit{170}$Er} & $^{201}$Ta & $^{218}$Ir & $^{233}$Hg \\
$^{82}$Br & $^{102}$Tc & $^{137}$Sb & $^{154}$Pm & $^{172}$Er & $^{204}$Ta & $^{219}$Ir &  \\
$^{83}$Br & $^{103}$Tc & \textit{$^\textit{126}$Te} & $^{155}$Pm & $^{184}$Er & $^{205}$Ta & $^{220}$Ir &  \\
$^{85}$Br & $^{105}$Tc & \textit{$^\textit{128}$Te} & $^{156}$Pm & $^{185}$Er & $^{207}$Ta & $^{221}$Ir &  \\
$^{87}$Br & \textit{$^\textit{102}$Ru} & \textit{$^\textit{130}$Te} & $^{157}$Pm & $^{186}$Er & $^{209}$Ta & $^{223}$Ir &  \\
\textit{$^\textit{81}$Kr} & \textit{$^\textit{104}$Ru} & $^{131}$Te & $^{158}$Pm & $^{194}$Er & $^{188}$W & $^{225}$Ir &  \\
\textit{$^\textit{82}$Kr} & $^{105}$Ru & $^{132}$Te & $^{159}$Pm & $^{196}$Er & $^{189}$W & \textit{$^\textit{196}$Pt} & 
\end{tabular}
\end{ruledtabular}
\end{table}
\endgroup

\begin{table}
\caption{\label{tab:ap} Targets for ($\alpha$,p) reactions with negative $Q$ value but smaller SEF than their inverse reaction. Stable or longlived targets are in italics.}
\begin{ruledtabular}
\begin{tabular}{cccccccc}
$^{39}$Ca & $^{66}$Ga & $^{72}$Kr & $^{86}$Zr & $^{97}$Pd & $^{101}$Sb & $^{111}$Cs & $^{134}$Gd \\
\textit{$^\textit{41}$Ca} & $^{62}$Ge & $^{74}$Kr & $^{87}$Zr & $^{97}$Ag & $^{105}$Sb & $^{120}$Cs & $^{139}$Tb \\
$^{56}$Ni & $^{64}$Ge & $^{76}$Kr & $^{84}$Mo & $^{98}$Ag & $^{107}$Sb & $^{116}$Ce & $^{138}$Dy \\
$^{57}$Ni & $^{65}$Ge & $^{78}$Rb & $^{86}$Mo & $^{98}$Cd & $^{108}$Sb & $^{118}$Ce & $^{145}$Tm \\
$^{58}$Cu & $^{66}$Ge & $^{76}$Sr & $^{88}$Mo & $^{100}$Cd & $^{109}$Sb & $^{120}$Ce & $^{146}$Tm \\
$^{60}$Cu & $^{70}$As & $^{80}$Sr & $^{89}$Mo & $^{103}$Cd & $^{101}$Te & $^{125}$Pr & $^{147}$Tm \\
$^{58}$Zn & $^{68}$Se & $^{81}$Sr & $^{91}$Mo & $^{99}$In & $^{102}$Te & $^{127}$Pr &  \\
$^{59}$Zn & $^{69}$Se & $^{83}$Sr & $^{92}$Ru & $^{103}$In & $^{104}$Te & $^{124}$Nd &  \\
$^{60}$Zn & $^{70}$Se & $^{80}$Zr & $^{93}$Ru & $^{104}$In & $^{106}$Te & $^{126}$Nd &  \\
$^{61}$Zn & $^{71}$Se & $^{82}$Zr & $^{94}$Ru & $^{97}$Sn & $^{111}$I & $^{129}$Pm &  \\
$^{62}$Zn & $^{73}$Se & $^{83}$Zr & $^{95}$Pd & $^{99}$Sn & $^{106}$Xe & $^{130}$Sm &  \\
$^{64}$Ga & $^{74}$Br & $^{84}$Zr & $^{96}$Pd & $^{100}$Sn & $^{112}$Xe & $^{135}$Eu & 
\end{tabular}
\end{ruledtabular}
\end{table}

Tables \ref{tab:ag}$-$\ref{tab:pn} list the reactions found to have $f_\mathrm{rev}<f_\mathrm{forw}$ according to the criteria discussed above. Stable or longlived target nuclei are marked specifically. Figures \ref{fig:gamma} and \ref{fig:particle} locate the reactions in the nuclear chart.

To current knowledge, not all shown reactions are of astrophysical importance. Others may be too far from stability to be accessible to experiments. Figure \ref{fig:gamma} displays capture or photodisintegration reactions. Of particular importance among them are the ($\alpha$,$\gamma$) reactions, also given in Table \ref{tab:ag}. They are located in a mass region which is of interest in the $p$- or $\gamma$-process \cite{woo78,arn03}. Although the $\gamma$ process synthesizes nuclei via photodisintegrations, it becomes obvious that even for $\alpha$ captures with negative $Q$ value the SEF of the capture reaction is smaller than the one of the photodisintegration. Thus, a measurement of the capture includes more astrophysically relevant transitions and is closer to the stellar value than a measurement of the photodisintegration in the laboratory. Many of the cases with negative $Q$ value are found at stability, providing interesting examples for experimental study. A second class of interesting reactions are the (p,$\gamma$) and ($\gamma$,p) reactions in Fig.\ \ref{fig:gamma} and Tables \ref{tab:pg}, \ref{tab:gpnang}. They involve unstable targets and are of interest in the $rp$ process \cite{sch98} and the $\nu p$ process \cite{froh}.

Figure \ref{fig:particle} displays reactions without a photon channel. Mostly interesting is the large number of (p,n) reactions located along stability (see Table \ref{tab:pn}). A recent investigation of the $\gamma$ process has pointed out the importance of (n,p) reactions close to stability \cite{rap06}. According to our findings, it is best to study the endothermic (p,n) reaction when trying to experimentally constrain the astrophysical rate. Our measurement of $^{85}$Rb(p,n)$^{85}$Sr, described in the following section, is an example for such an experiment.

Of minor or no astrophysical relevance are the ($\alpha$,n) and (n,p) reactions in Fig.\ \ref{fig:particle} and Tables \ref{tab:an}, \ref{tab:np} because they involve neutron-rich targets and these reactions are much slower than other possible reactions on the same nuclei.

Figure \ref{fig:gamma} and Tables \ref{tab:gn}, \ref{tab:gpnang} list (n,$\gamma$) and ($\gamma$,n) reactions on very neutron-rich nuclei far off stability. They are relevant in the $r$ process \cite{cow91,arngor}. Obviously, the suppression of the SEF is not caused by the Coulomb barrier. The nuclear structure (spins and parities of excited states and the nuclear level density) is important in those cases, as the level density is low and transitions are favored or suppressed by selection rules and centrifugal barriers. Contrary to the Coulomb suppression of the SEF, this type of suppression is strongly dependent on the nuclear spectroscopy assumed in the calculation. Moreover, the Hauser-Feshbach model of compound reactions may not be applicable anymore for the nuclei involved \cite{rtk97}. Additionally, the individual rates are not important in $r$ process studies producing highly unstable, neutron-rich nuclei close to the dripline in an equilibrium between capture and photodisintegration \cite{frei99,faro09}.

\section{Experimental study of $^{85}$R\lowercase{b(p,n)}$^{85}$S\lowercase{r}}

\subsection{General}
\label{sec:exp}

As an example of the
suppression effect and for the derivation of the astrophysical rates for an endothermic reaction, we experimentally studied the reaction $^{85}$Rb(p,n)$^{85}$Sr.

Reactions of the (n,p) type have been shown to be important in the $\gamma$ process \cite{rap06}. This nucleosynthesis process creates proton-rich isotopes of elements beyond Fe which are not made in the $s$ and $r$ processes. It was shown to occur in hot O/Ne layers of massive stars, either in a core collapse supernova explosion when the shockfront is passing these layers or already pre-explosively depending on the initial mass of the star \cite{rau02}. At temperatures $T>2$ GK photodisintegrations can act even within the short timescale of an explosion.
The reaction sequences initially drive material from the bottom of the valley
of stability to the proton-rich side by ($\gamma$,n) reactions.
Charged-particle emitting ($\gamma$,$\alpha$) and ($\gamma$,p) reactions
can deflect the reaction path to lower charge number.
Theoretical investigations show that ($\gamma$,n)/($\gamma$,p) branchings
play a key role in the production of the lighter $p$ nuclei
whereas ($\gamma$,n)/($\gamma$,$\alpha$) branchings are important at higher masses \cite{rau06,rap06}.
Further reactions with the emitted neutrons are mainly important in the freeze-out phase when photodisintegration ceases \cite{rap06,dill08}. The flow back to stability is sped up by (n,p) reactions which are faster than $\beta$ decays close to stability \cite{rap06}. Even at stability, (n,p) reactions act to push material to lower proton numbers. In this context, $^{85}$Rb(p,n)$^{85}$Sr is directly important because it is the inverse reaction to $^{85}$Sr(n,p) and we found that its SEF is smaller than the one of its inverse, despite of its negative $Q$ value (see Sec.\ \ref{sec:astrorate}).

The reaction $^{85}$Rb(p,n)$^{85}$Sr is also important to test the predictions of astrophysical rates and their underlying nuclear properties. Although many (n,p) and (p,$\gamma$) reactions important in the $\gamma$, $rp$, and $\nu p$ processes \cite{rap06,sch98,froh} occur far from stability, the models and assumptions used in the prediction of the rates can be checked at stability. Especially suited for testing the reliability of the optical potential used for the calculation of transitions involving protons are (n,p) and (p,n) reactions because the proton width is smaller than the neutron width at practically all energies (except very close to the neutron threshold) and thus determines the cross section.

We measured $^{85}$Rb(p,n)$^{85}$Sr
using the activation method. Thin RbCl targets were bombarded by proton beam provided by the Van de Graaff and cyclotron accelerators of ATOMKI \cite{kiss08}. 
The (p,n) reaction on $^{85}$Rb can populate both the ground and metastable states of $^{85}$Sr \cite{decay}. To determine the cross section of the $^{85}$Rb(p,n)$^{85}$Sr$^g$ reaction the 514.01 keV $\gamma$ line was used, in the case of the $^{85}$Rb(p,n)$^{85}$Sr$^m$ reaction cross section the yield of the 231.84 keV transition was measured. In the following Secs.\ \ref{sec:exptarget}--\ref{sec:activity} a detailed description of the experimental technique is given, while the experimental results are given in Sec.\ \ref{sec:expresults}. A comparison to theory and the final astrophysical reaction rates are provided in Secs.\ \ref{sec:potential} and \ref{sec:astrorate}.

\begin{figure}
\resizebox{0,7\columnwidth}{!}{\rotatebox{270}{\includegraphics[clip=]{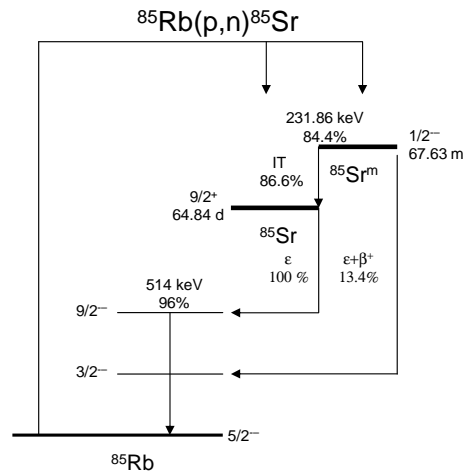}}}
\caption{\label{fig:decay} Simplified decay scheme of the products of the $^{85}$Rb(p,n)$^{85}$Sr reaction. The half-lives of the reaction products, the branching ratios, the spin and parity of the levels and the transitions used to determine the reaction cross section are indicated \cite{decay}. }
\end{figure}

\begin{figure*}
\resizebox{1.4\columnwidth}{!}{\rotatebox{270}{\includegraphics[clip=]{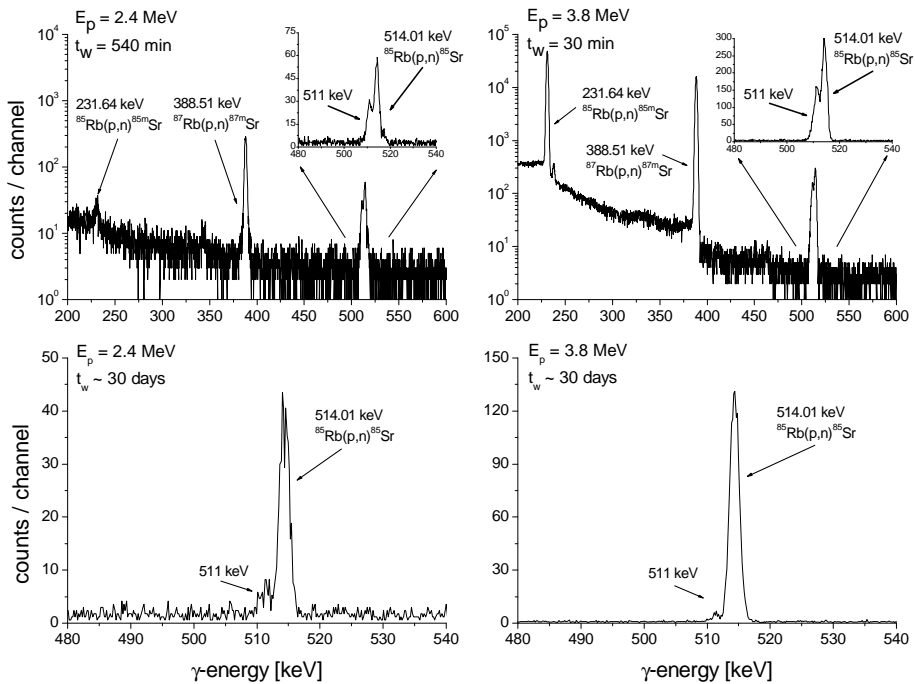}}}
\caption{\label{fig:spectra}Typical $\gamma$-spectra taken after the irradiation of RbCl targets with 2.4 (left panel) and 3.8 MeV (right panel) proton beams. The 514.01 keV peak from the $^{85}$Rb(p,n)$^{85}$Sr$^g$ reaction can be well separated from the annihilation peak as can be seen in the insets. The length of the waiting time (t$_w$) between the end of the irradiation and the start of the $\gamma$-countings were 540 (E$_p$ = 2.4 MeV) and 30 min (E$_p$ = 3.8 MeV). The lower panels show typical spectra taken in the repeated activity measurement approximately one month after the irradiations (for details see text).}
\end{figure*}

\subsection{Target properties and the determination of the number of target atoms}
\label{sec:exptarget}

The targets were made by evaporating chemically pure (99.99\%) RbCl onto two different kinds of Aluminum foils: the thicker one had a chemical purity of 99.999\% and thickness of 50 $\mu$m, while the purity and the thickness of the thinner one was 99\% and 2.4 $\mu$m, respectively. The distance between the evaporation boat and the target backing was 10 cm, therefore it was possible to assume that the evaporated layer is homogeneous. This assumption was proved using Rutherford Backscattering Spectroscopy (RBS, see later). Targets with different thicknesses were used, thicker ones (on thicker backings) were employed for irradiations at lower and thinner ones (on thinner backing) at higher energy. Owning to this treatment, the yield of the investigated 514.01 keV peak was always higher than, or comparable to that of the 511 keV annihilation peak -- and this way the separation of the peaks was achieved -- as it is demonstrated in the upper part of Fig.\ \ref{fig:spectra}.

The number of the target atoms was determined with Rutherford Backscattering Spectrometry (RBS) at the Nuclear Microprobe facility of ATOMKI \cite{raj96,sim06, yal09}. As a consistency check, in the case of the targets evaporated onto the thinner backing, weighing was also used to determine the number of target atoms. The weight of the Al foil used as backing was measured before and after the evaporation and from the difference --- assuming that our target is uniform --- the number of target atoms was calculated. The results of the two different methods used to determine the number of target atoms are in very good agreement ($\leq$ 3\% difference).

\subsection{Irradiation}
\label{sec:irr}

The RbCl targets were bombarded with a proton beam provided by the Van de Graaff and cyclotron accelerators of ATOMKI. The energy of the proton beam was between 2 and 4 MeV, this energy range was covered in 200 keV steps. The beam current was typically 600 nA. Each irradiation lasted approximately $7-8$ hours. The low energy irradiations (2.2 MeV, 2.4 MeV and 2.6 MeV) have been carried out using the Van de Graaff. At and above 2.6 MeV the cyclotron accelerator was used. The cross section at E$_p$ = 2.6 MeV was measured with both accelerators and no difference was found. 

An ion implanted Si detector was built into the irradiation chamber at $\theta$ = 150$^{\circ}$ relative to the beam direction to measure the yield of the backscattered protons. The backscattering spectra were taken continuously and were used to monitor the target stability. Having a beam restricted to 600 nA, no target deterioration was found.

For calculating the reaction cross section the proper knowledge of the incident particle flux is necessary. To obtain this, the collected charge was measured in a chamber similar to the one in \cite{kiss07}. After the beam defining aperture, the whole chamber served as Faraday cup to collect the accumulated charge. A secondary electron suppression voltage of $-300$ V was applied at the entrance of the chamber. The beam current was kept as stable as possible but to follow the changes the current integrator counts were recorded in multichannel scaling mode, stepping the channel in every minute. This recorded current integrator spectrum was used for the analysis solving the differential equation of the population and decay of the reaction products numerically. 

\subsection{Activity determination}
\label{sec:activity}

Figure \ref{fig:decay} shows the simplified decay scheme of the $^{85}$Sr$^{g,m}$ isotopes. To determine the cross section of the $^{85}$Rb(p,n)$^{85}$Sr$^g$ reaction the 514.01 keV, for the $^{85}$Rb(p,n)$^{85}$Sr$^m$ reaction the 231.84 keV gamma line was used. 

For measuring the induced $\gamma$-activity a lead shielded HPGe detector was used as in our previous (p,n)-study \cite{kiss07}. After each irradiation, a cooling time of one hour was inserted in order to let the disturbing shortlived activities decay. The $\gamma$ spectra were taken for 12 hours and stored regularly in order to follow the decay of the short-lived reaction product.

Figure \ref{fig:spectra} shows typical spectra collected after irradiating RbCl targets with the 2.4 MeV (left panel) and the 3.8 MeV (right panel) proton beams.
The yield of the 511 keV peak was always less than or comparable to the investigated 514.01 keV transition, as shown in the insets. The $^{85\mathrm{g}}$Sr has a relatively long halflife ($T_{1/2}$ = 64.84 d). Due to this, the activity measurement could be repeated for each target after approximately one month, when the intensity of the 511\,keV radiation was substantially reduced. The spectra taken in the repeated activity measurement for the 2.4 and 3.8 MeV irradiations are shown in the lower panels of Fig.\ \ref{fig:spectra}. The two measurements yielded consistent cross sections proving the proper separation of the 511\,keV and 514.01\,keV peaks.

\begin{table*}
\caption{\label{tab:exp_results}Details of the irradiations and the resulted cross sections (astrophysical $S$ factors).}
\begin{tabular}{ccccccc}
\parbox[t]{1.4cm}{\centering{E$_{lab.}$} \\ $\left[ MeV\right]$} &
\parbox[t]{1.24cm}{\centering{E$_{c.m.}$} \\ $\left[ MeV\right]$} &
\parbox[t]{2.8cm}{\centering{Accelerator}} &
\parbox[t]{1.8cm}{\centering{Collected \\ charge $\left[ mC\right]$}} &
\parbox[t]{2.8cm}{\centering{Total $\sigma$ \\ $\left[ mbarn\right]$}} &
\parbox[t]{2.8cm}{\centering{ $S$ factor \\ $\left[ 10^6 MeV barn\right]$ }} \\
\hline\hline
2.20  & 2.16 $\pm$ 0.008  & Van de Graaff &        48.27    & 0.058  $\pm$ 0.006   & 7.13  $\pm$ 0.67 \\
2.40  & 2.34 $\pm$ 0.016  & Van de Graaff &        37.23    & 0.224  $\pm$ 0.019   & 11.22 $\pm$ 0.96 \\
2.60  & 2.57 $\pm$ 0.026  & Van de Graaff &        33.74    & 0.582  $\pm$ 0.055   & 11.01 $\pm$ 1.04 \\
2.58  & 2.55 $\pm$ 0.027  & Cyclotron     &        35.44    & 0.569  $\pm$ 0.051   & 11.35 $\pm$ 1.02 \\
2.79  & 2.77 $\pm$ 0.028  & Cyclotron     &        30.83    & 1.20   $\pm$ 0.12    & 10.65 $\pm$ 1.05 \\
2.98  & 2.96 $\pm$ 0.030  & Cyclotron     &        23.66    & 2.12   $\pm$ 0.21    & 9.59  $\pm$ 0.92 \\
3.18  & 3.16 $\pm$ 0.032  & Cyclotron     &        23.43    & 3.77   $\pm$ 0.35    & 9.40  $\pm$ 0.87 \\
3.37  & 3.34 $\pm$ 0.035  & Cyclotron     &        21.45    & 5.66   $\pm$ 0.54    & 8.37  $\pm$ 0.79 \\
3.57  & 3.55 $\pm$ 0.036  & Cyclotron     &        20.64    & 9.60   $\pm$ 0.87    & 8.32  $\pm$ 0.76 \\
3.76  & 3.75 $\pm$ 0.037  & Cyclotron     &        17.16    & 14.31  $\pm$ 1.22    & 7.73  $\pm$ 0.66 \\
3.96  & 3.95 $\pm$ 0.040  & Cyclotron     &        11.51    & 19.65  $\pm$ 1.82    & 6.93  $\pm$ 0.64 \\
\hline
\end{tabular}
\end{table*}

\subsection{Experimental results and comparison with literature data}
\label{sec:expresults}

In the case of the $^{85}$Rb(p,n)$^{85}$Sr$^{g}$ reaction, two separated analysis were done. The agreement between the cross sections derived in the $\gamma$-counting after the irradiation and the ones from the repeated activity measurement was always better than 4\%. The final results were calculated from the average weighted by the statistical uncertainty of the two $\gamma$ countings. The halflife of the $^{85}$Sr$^{m}$ is shorter, therefore the yield of the 231.64 keV $\gamma$ radiaton was measured only after the irradiation. The final experimental result can be found in Table \ref{tab:exp_results}. Partial cross sections leading to the ground and isomeric state of $^{85}$Sr can be found in \cite{kiss08}.
The error of the cross section values is the quadratic sum of the following partial errors: efficiency of the HPGE detector system (6\%), number of target atoms ($\leq$ 3.3\%), current measurement (3\%), uncertainty of the level parameters found in the literature ($\leq$ 4\%) and counting statistics ($\leq$ 4\%). The quoted errors of the energies include the energy loss in the targets calculated with the SRIM code \cite{SRIM}, as well as the energy stability of the cyclotron and Van de Graaff accelerators.

The measured total cross sections cover 3 orders of magnitude, varying from 0.06 to 20 mb.
Table \ref{tab:exp_results} lists the measured cross sections $\sigma$ and the $S$ factors, the latter being defined as \cite{ilibook}
\begin{equation}
S(E)=\frac{\sigma}{E} e^{-2\pi \eta} \quad,
\end{equation}
with the Sommerfeld parameter $\eta$ accounting for the Coulomb barrier penetration.

The cross section of the $^{85}$Rb(p,n)$^{85}$Sr reaction was already investigated by \cite{kastleiner02} between E$_{c.m.}$ = 3.1 and 70.6 MeV. However, their accuracy is not sufficient for astrophysical applications, mainly because of the large uncertainty of the c.m.\ energies. Moreover, there is only one data point in the relevant energy region for the $\gamma$ process and it bears an uncertainty of $\pm$ 0.5 MeV in the c.m.\ energy. 

\subsection{Comparison to theory and implications for the proton optical potential}
\label{sec:potential}

\begin{figure}
\resizebox{\columnwidth}{!}{\rotatebox{270}{\includegraphics[clip=]{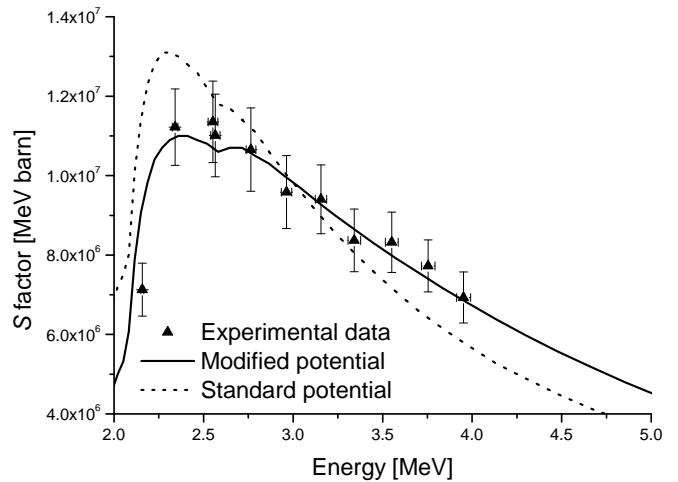}}}
\caption{\label{fig:sfact}Experimental (full triangles) and theoretical (lines) astrophysical $S$ factors of $^{85}$Rb(p,n)$^{85}$Sr.
The solid line is the $S$ factor calculated with the modified proton
optical potential introduced in \cite{kiss07} and the dashed line shows the
result using the standard proton optical potential from \cite{jeu77} with low
energy modifications by \cite{lej80} (see text).}
\end{figure}

The measured $S$ factors are compared to theoretical predictions obtained with the code NON-SMOKER 
\cite{nonsmoker,adndt}
in Fig.\ \ref{fig:sfact}. The standard calculation applied a proton optical potential widely
used in astrophysical applications, based on a microscopic approach utilizing a local density approximation \cite{jeu77}.
Low-energy modifications, which are relevant in astrophysics, have been provided by \cite{lej80}. As can be seen in Fig.\ \ref{fig:sfact}, the theoretical energy dependence of the resulting $S$ factor is slightly steeper than the data, although
there is general agreement in magnitude. In the energy range covered by the measurement, the proton width is smaller
than the neutron width (except close to the threshold) and thus uncertainties in the description of the proton width
(and proton transmission coefficient) will fully impact the resulting $S$ factor.
A recent investigation \cite{kiss07} suggested that the strength of the
imaginary part of the microscopic potential should be increased by 70\%.
We find that the energy-dependence of
the theoretical $S$ factor is changed in such a way as to show perfect agreement with the new data,
as seen in Fig.\ \ref{fig:sfact}.
This independently confirms the conclusions of previous work \cite{kiss07}.

\subsection{Astrophysical reaction rates}
\label{sec:astrorate}

Regarding the Coulomb suppression effect, a comparison of $1.03\leq f_\mathrm{pn} \leq 1.08$
and $2.6\leq f_\mathrm{np} \leq 3.9$ shows that the transitions to
excited states of $^{85}$Sr are more important than those to states in $^{85}$Rb
in the relevant plasma temperature range of $2 \leq T \leq 4$ GK.
The almost negligible stellar enhancement $f_\mathrm{pn}$ is due to the suppression of
the proton transmission coefficients
to and from the excited states of $^{85}$Rb
for small relative proton energies because of the Coulomb barrier. There are only few
transitions able to contribute due to the low $Q$ value.
As shown by the small $f_\mathrm{pn}$, the transition from the ground state of $^{85}$Rb dominates the proton channel.
Obviously, a Coulomb suppression is not present in the neutron channel.
On the contrary, for this reaction $f_\mathrm{np}$ is even more enhanced
due to the spin structure of the available nuclear levels and especially the large spin of $^{85}$Sr$^g$.
Because of its large spin, it is connected to the (dominating) low spin states in $^{85}$Rb
through higher partial waves
than the excited states, such as the isomeric state, which have lower spins. Thus, the transitions from the ground state are suppressed by the centrifugal barrier relative to transitions from excited states and the latter will quickly become important, even at low temperature. As a consequence of the enhancement of $f_\mathrm{np}$ and the suppression of $f_\mathrm{pn}$, it is more advantageous to measure the (p,n) direction. Important transitions to states
in $^{85}$Sr are included in our data and the small impact of transitions from excited states in $^{85}$Rb is within the experimental error.

\begin{table}
\caption{\label{tab:rates}Astrophysical reactivities $N_\mathrm{A}\mathcal{R}^*$ of the reactions
$^{85}$Rb(p,n)$^{85}$Sr and $^{85}$Sr(n,p)$^{85}$Rb computed from
experimental data. The values in italics are at temperatures where the experimental data mostly contribute to the rate.
The other values are computed by supplementing theoretical cross sections using the modified optical potential.}
\begin{ruledtabular}
\begin{tabular}{lr@{$\pm$}lr@{$\pm$}l}
\multicolumn{1}{c}{Temperature}&\multicolumn{2}{c}{$^{85}$Rb(p,n)$^{85}$Sr}&\multicolumn{2}{c}{$^{85}$Sr(n,p)$^{85}$Rb} \\
\multicolumn{1}{c}{[$10^9$ K]}&\multicolumn{2}{c}{[cm$^3$s$^{-1}$mole$^{-1}$]}&\multicolumn{2}{c}{[cm$^3$s$^{-1}$mole$^{-1}$]}\\
\hline
  0.10 & (1.72&0.17)$\times 10^{-89}$ & (1.19&0.2)$\times 10^{4}$ \\  
  0.15 & (2.21&0.22)$\times 10^{-58}$ & (1.49&0.15)$\times 10^{4}$ \\  
  0.20 & (8.33&0.83)$\times 10^{-43}$ & (1.74&0.17)$\times 10^{4}$ \\  
  0.30 & (3.36&0.33)$\times 10^{-27}$ & (2.15&0.22)$\times 10^{4}$ \\  
  0.40 & (2.26&0.23)$\times 10^{-19}$ & (2.55&0.26)$\times 10^{4}$ \\  
  0.50 & (1.18&0.12)$\times 10^{-14}$ & (2.99&0.30)$\times 10^{4}$ \\  
  0.60 & (1.74&0.17)$\times 10^{-11}$ & (3.49&0.35)$\times 10^{4}$ \\  
  0.70 & (3.31&0.33)$\times 10^{-9}$ & (4.09&0.41)$\times 10^{4}$ \\  
  0.80 & (1.77&0.18)$\times 10^{-7}$ & (4.80&0.48)$\times 10^{4}$ \\  
  0.90 & (4.04&0.40)$\times 10^{-6}$ & (5.62&0.56)$\times 10^{4}$ \\  
  1.00 & (5.07&0.51)$\times 10^{-5}$ & (6.57&0.66)$\times 10^{4}$ \\  
  1.50 & (1.28&0.13)$\times 10^{-1}$ & (1.35&0.14)$\times 10^{5}$ \\ 
{\it  2.00 } & {\it (8.30}&{\it 0.83)} & {\it (2.56}&{\it 0.26)}$\mathit{\times 10^{5}}$ \\  
{\it  2.50 } & {\it (1.21}&{\it 0.12)}$\mathit{\times 10^{2}}$ & {\it (4.57}&{\it 0.46)}$\mathit{\times 10^{5}}$ \\  
{\it  3.00 } & {\it (8.22}&{\it 0.82)}$\mathit{\times 10^{2}}$ & {\it (7.81}&{\it 0.78)}$\mathit{\times 10^{5}}$ \\  
{\it  3.50 } & {\it (3.56}&{\it 0.36)}$\mathit{\times 10^{3}}$ & {\it (1.28}&{\it 0.13)}$\mathit{\times 10^{6}}$ \\  
{\it  4.00 } & {\it (1.15}&{\it 0.12)}$\mathit{\times 10^{4}}$ & {\it (2.04}&{\it 0.20)}$\mathit{\times 10^{6}}$ \\  
  4.50 & (3.03&0.30)$\times 10^{4}$ & (3.17&0.32)$\times 10^{6}$ \\  
  5.00 & (6.89&0.69)$\times 10^{4}$ & (4.76&0.48)$\times 10^{6}$ \\  
  6.00 & (2.60&0.26)$\times 10^{5}$ & (9.52&0.95)$\times 10^{6}$ \\  
  7.00 & (7.14&0.71)$\times 10^{5}$ & (1.54&0.15)$\times 10^{7}$ \\  
  8.00 & (1.50&0.15)$\times 10^{6}$ & (2.01&0.20)$\times 10^{7}$ \\  
  9.00 & (2.50&0.25)$\times 10^{6}$ & (2.18&0.22)$\times 10^{7}$ \\  
 10.00 & (3.44&0.34)$\times 10^{6}$ & (2.05&0.21)$\times 10^{7}$
\end{tabular}
\end{ruledtabular}
\end{table}

Applying Eq.\ \ref{eq:rate} directly with the experimental cross sections already yields the stellar rate because the SEF is small in the (p,n) direction. The stellar rate of the exothermic (n,p) reaction can then be computed using Eq.\ (\ref{eq:revrate}).
By computing the forward rates directly from the backward rates without using fits,
the complication with the negative $Q$ value in fitted data is also avoided.

Table \ref{tab:rates} gives the stellar reactivities (as defined by Eq.\ \ref{eq:reactivity}) for $^{85}$Rb(p,n)$^{85}$Sr as well as for $^{85}$Sr(n,p)$^{85}$Rb. Our data
covers an energy range sufficient to compute the rates between 2 and 4 GK. Because of the excellent agreement of theory
with experiment (using the newly modified potential of \cite{kiss07}), we supplement the data with the theoretical values to compute the reactivities at lower and higher temperatures, applying
the same errors as for the data.
\begin{figure}
\resizebox{\columnwidth}{!}{\rotatebox{270}{\includegraphics[clip=]{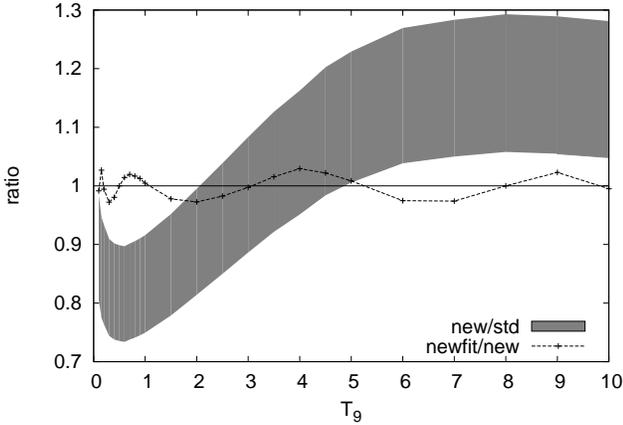}}}
\caption{The newly derived stellar reactivity of $^{85}$Sr(n,p)$^{85}$Rb is compared to the one given in \cite{adndt01} (area labelled ``new/std''). The temperatures $T_9$ are stellar plasma temperatures in GK. The shaded area accounts for an error of $\pm 10$\%. As seen in Table \ref{tab:rates}, the experimental data contribute significantly in the range $2\leq T_9\leq 4$. Also shown is the reactivity from a fit of our new result compared to the actual value (curve labelled ``newfit/new''). This shows that the fit accuracy is high.\label{fig:ratios}
}
\end{figure}

It is to be noted that fits of the rates should be obtained by first fitting the (n,p)
rate and then deriving the (p,n) rate fit by modifying the fit coefficients according to detailed balance as given in Eq.\ (\ref{eq:revrate}) (see \cite{adndt} for details). For convenience, we provide the fit coefficients (including a 10\% error) for the (n,p) reactivity in the widely used REACLIB format \cite{adndt,cow91}
\begin{eqnarray}
N_\mathrm{A}\mathcal{R}^*&=&\exp \left( a_0+\frac{a_1}{T_9}+\frac{a_2}{T_9^{1/3}}+a_3T_9^{1/3}+a_4T_9 \right. \nonumber \\
&&\left. +a_5T_9^{5/3}+a_6\ln T_9 \right) \quad,
\end{eqnarray}

where $N_\mathrm{A}$ is Avogadro's number and the plasma temperature $T_9=T/10^9$, with $T$ in K.
Using the usual dimension of cm$^3$ s$^{-1}$ mole$^{-1}$ for $N_\mathrm{A}\mathcal{R}^*$ the fitted coefficients evaluate to $a_0=33.2271^{+\ln{1.1}}_{+\ln{0.9}}$, $a_1=-0.886129$, $a_2=40.7296$, $a_3=-67.9553$, $a_4=6.54471$, $a_5=-0.562194$, $a_6=31.1997$. The assumed error is contained in the error given for $a_0$. The coefficients
for the (p,n) direction are the same, except $a_0^\mathrm{pn}=33.44895^{+\ln{1.1}}_{+\ln{0.9}}$ and $a_1^\mathrm{pn}=-22.3196405$. According to Eqs.\ (\ref{eq:reactivity}) and (\ref{eq:revrate}), to obtain the final (p,n) rate the value obtained with the seven parameter expression has to be multiplied not only by the number densities of the interacting particles but also by the ratio of the temperature-dependent partition functions of initial and final nucleus
\begin{eqnarray}
N_\mathrm{A}\mathcal{R}'_\mathrm{pn}(T)&=&\exp \left( a_0^\mathrm{pn}+\frac{a_1^\mathrm{pn}}{T_9}+\frac{a_2}{T_9^{1/3}}+a_3T_9^{1/3}+a_4T_9 \right. \nonumber \\
&&\left. +a_5T_9^{5/3}+a_6\ln T_9 \right) \quad, \\
N_\mathrm{A}\mathcal{R}^*_\mathrm{pn}(T)&=&\frac{G_{^{85}\mathrm{Sr}}(T)}{G_{^{85}\mathrm{Rb}}(T)}N_\mathrm{A}\mathcal{R}'_\mathrm{pn}(T) \quad .
\end{eqnarray}
The required partition functions $G(T)$ are provided in \cite{adndt} as a function of temperature.

Figure \ref{fig:ratios} shows a comparison of the new $^{85}$Sr(n,p)$^{85}$Rb reactivity to the ``standard'' one of \cite{adndt01}. At temperatures above 3 GK, we see an increase of $10-30$\% compared to the previous values. Below 2 GK, the new reactivity is $20-30$\% lower than previously. The change in the temperature dependence is due to the different proton optical potential used. At very low temperature, the reactivity becomes less sensitive to the proton potential. This explains the ratio becoming almost unity towards zero temperature. Also shown in Fig.\ \ref{fig:ratios} is a comparison between the fit of the new reactivity with the parameters above and the reactivity itself. This ratio stays close to unity for all temperatures. The deviations between the reactivity and its fit are very small and negligible compared to the other uncertainties involved.

\section{Summary}
\label{sec:summary}

We showed that -- contrary to common wisdom -- a large number of endothermic reactions exhibit smaller stellar enhancement than their exothermic counterparts and are thus preferable for experimental studies. The main cause of suppression of the SEF in an endothermic reaction is the Coulomb suppression of transitions with low relative energy. This Coulomb suppression of the SEF was found to act for
reactions with $Q<0$ but low $\left| Q\right|$ and charged projectiles. Allowing only nucleons,
$\alpha$ particles, and photons as projectiles or ejectiles, and restricting the results to experimentally useful values of the SEFs, this effect
still appears in 1200 reactions, including $\alpha$ captures
relevant in the $p$ process \cite{rau06,rap06} and proton captures relevant
in the $rp$ process \cite{sch98} and the $\nu p$ process \cite{froh}. A large number of cases was also found for (p,n) reactions which allow to determine astrophysical reaction rates relevant to the $\gamma$ process \cite{rap06}.

As an example, we measured the astrophysically important
reaction $^{85}$Rb(p,n)$^{85}$Sr close above the threshold in the energy range relevant for the $\gamma$ process. It was shown that in this case it is possible to derive
astrophysical reaction rates for the (n,p) as well as the (p,n)
direction directly from the (p,n) data despite of the negative reaction $Q$ value.
Additionally, our measurement confirms a previously derived modification of the global proton optical potential used in theoretical predictions.

\acknowledgments
This work was supported by the European Research Council grant 
agreement no. 203175, the Economic Competitiveness 
Operative Programme GVOP-3.2.1.-2004-04-0402/3.0., OTKA (K68801, T49245),
and the Swiss NSF (grant 2000-105328).
Gy.\ Gy.\ acknowledges support from the Bolyai grant.

\end{document}